\documentclass[twocolumn]{revtex4}
\usepackage{graphicx}
\usepackage[]{epsfig}
\usepackage{epstopdf}
\newcommand{\beq}{\begin{equation}}
\newcommand{\eeq}{\end{equation}}
\newcommand{\beqd}{\begin{displaymath}}
\newcommand{\eeqd}{\end{displaymath}}
\newcommand{\beqa}{\begin{eqnarray}}
\newcommand{\eeqa}{\end{eqnarray}}

\newcommand{\comment}[1]{}
\newcommand{\Tr}{{\rm Tr}}

\begin{document}
\title{Universality and Deviations in Disordered Systems}


\author{Giorgio Parisi$^{1}$ and Tommaso Rizzo$^{2}$}

\affiliation{$^{1}$Dipartimento di Fisica, Universit\`a di Roma ``La Sapienza'', 
P.le Aldo Moro 2, 00185 Roma,  Italy
\\
$^{2}$ ``E. Fermi'' Center, Via Panisperna 89 A, Compendio Viminale, 00184, Roma, Italy}

\begin{abstract}
We compute the probability of positive large deviations of the free energy {\it per spin} in mean-field Spin-Glass models.
The probability vanishes in the thermodynamic limit as $P(\Delta f) \propto \exp[-N^2 L_2(\Delta f)]$.
For the Sherrington-Kirkpatrick model we find $L_2(\Delta f)=O(\Delta f)^{12/5}$ in good agreement with numerical data and with the assumption that typical small deviations of the free energy scale as $N^{1/6}$.
For the spherical model we find $L_2(\Delta f)=O(\Delta f)^{3}$ in agreement with recent findings on the fluctuations of the largest eigenvalue of random Gaussian matrices.
The computation is based on a loop expansion in replica space and the non-gaussian behaviour follows in both cases from the fact that the expansion is divergent at all orders. The factors of the leading order terms are obtained resumming appropriately the loop expansion and display universality, pointing to the existence of a single universal distribution describing the small deviations of any model in the full-Replica-Symmetry-Breaking class.
\end{abstract} 

\maketitle

The free energy density of a random system model fluctuates over the disorder with a distribution that in the thermodynamic limit becomes peaked around the typical value $f_{typ}$. The probability of observing $O(1)$ deviations (i.e. large deviations) from the typical value are exponentially small. In some simple systems (like the mean field spherical model for spin glasses) the probability of large deviations is highly asymmetric \cite{DM}.
In the case of negative deviations $\Delta f=f-f_{typ}$ we have $P(\Delta f) \propto \exp[-N L(\Delta f)]$ \cite{PR1};
for positive deviations the probability is much smaller:  $P(\Delta f) \propto \exp[-N^2 L_2(\Delta f)]$.  

This unusual asymmetric behaviour related to the distribution the large deviations of the lowest eigenvalue of a large Gaussian random matrix. The probability of observing large positive deviations scales indeed as $\exp[-O(N^2)]$ and had been recently computed in \cite{DM}. On the other hand the {\it typical} small deviation distribution of the lowest eigenvalue are described by a universal function in the large $N$-limit that
was computed by Tracy and Widom \cite{TW} and has since appeared in many apparently unrelated problems (see {\it e.g.} \cite{DM}).

In this letter we show that  in the mean-field  Sherrington-Kirkpatrick(SK) spin-glass model a similar scaling is present and the probability distribution $ L_2(\Delta f)$ can be computed  using the hierarchical replica-symmetry-breaking(RSB) ansatz (the function $L(\Delta f)$ that is relevant for negative deviations was computed in \cite{PR1}).
Interestingly enough our results also show a great deal of universality suggesting that the small deviations of the free energy of the whole class of {\it full}-RSB models are also described by the same universal function at any temperature in the spin-glass phase.

The computation of $L_2(\Delta f)$ is much more complex than that of $L(\Delta f)$: indeed the function $L(\Delta f)$ is infinite for positive $\Delta f$ \cite{DFM,TALA}. If we attack the problem in the replica framework it leads to an action of a matrix $\tilde{Q}_{ab}$ of size $n \times n$ where $n= \alpha N$ with negative $\alpha$.  Usually fluctuations around the saddle point are negligible (they give a contribution to the total free energy that is proportional to $n^2$) however they play a crucial role in this case  because the number 
of elements of the matrix $\tilde{Q}$ is $O(N^2)$. We have also to face the fact the the saddle point is marginal (i.e. there are zero modes that lead to divergent fluctuations) thus if we proceed naively we obtain a series divergent at all orders.
The marginality of mean-field theory is a general well-known feature of full-RSB spin-glass and it stands as the main difficulty to solve many of the open problems in the field. 
In particular it leads to non-trivial finite-size correction in the SK model for which no systematic computation scheme has been devised up to now \cite{ABMM}. 
Besides the fact that no theory can be considered satisfactory if it does not allow to include deviations and corrections, these non-trivial mean-field exponents have been argued to be relevant also for finite-dimensional quantities, notably the stiffness exponent \cite{AM0}. 
Furthermore we do not know how to treat loop corrections to the theory in the spin-glass phase below six dimensions \cite{DKT}.
Thus the need to tackle the marginal nature of the theory makes the computation as interesting as the problem itself.

In order to compute the so called {\em sample complexity}  $L_2(\Delta f) \equiv-  \ln(P(\Delta f))/N^2$ we consider the average partition function of $n=\alpha N$ replicas: 
\beq
\Phi(\alpha) \equiv -{1 \over N^2}\ln \overline{Z^{\alpha N}}\, ,
\label{phia}
\eeq
with $\alpha$ negative (the bar denotes the average over the disorder). In the usual approach, when we need to compute the typical free energy ($f_{typ}$) the number ($n$) of replicas goes to zero, but in this case it has to go to $-\infty$.

The functional $\Phi(\alpha)$ is the Legendre transform of $L_2 (\Delta f)$. The latter can be obtained as 
$-L_2 (\Delta f)=\alpha \beta f_{typ}+\alpha \beta \Delta f- \Phi(\alpha)$ where $\alpha$ is the solution of $\beta f_{typ}+\beta \Delta f=d \Phi(\alpha)/d\alpha$. The main result of this letter is the evaluation of  $\Phi(\alpha)$ for small $\alpha$.

Let us now consider what happens in the SK spin-glass model where $H=\sum_{i,k}J_{i,k}\sigma_i\sigma_k$ with the $\sigma$'s being Ising spins and the $J$'s being random Gaussian variable with zero average and variance $1/N$.
Through standard manipulations we rewrite (\ref{phia}) as an the integral of an action depending on a symmetric matrix $\tilde{Q}_{ab}$ with $a,b=1,\dots,\alpha N$ and $\tilde{Q}_{aa}=0$.
As the size of the matrix is $O(N)$ we cannot simply take the saddle point and we have consider the integral of $O(N^2)$ elements of $\tilde{Q}_{ab}$. 
In order to perform the integration we find convenient to divide the matrix $\tilde{Q}_{ab}$ in $\left( {N \alpha \over n}\right)^2$ blocks of size $n \times n$, with $n$ some parameter between $1$ and $\alpha N$ that will be eventually sent to zero. The generic matrix element will be written as $Q_{ab}^{ij}$ where the upper indices $i,j=1,\dots,N\alpha/n$ label different blocks and the lower indices $a,b=1,\dots,n$ label elements inside block $ij$.

The resulting action admits a saddle point with vanishing off-diagonal blocks $Q^{ij}=0$ for all $i \neq j$ and 
 we integrate out the elements of the off-diagonal blocks $i\neq j$ around their zero saddle-point value. 
After the integration we are left with an integral over the diagonal blocks of the exponential of a $O(N^2)$ action over which we will take the saddle point. We will consider here the simplest situation where the  saddle points are such that all the blocks on the diagonal are equal to a given $n \times n$ hierarchical matrix $Q$, {\it i.e.} $Q^{ii}=Q$ $\forall i$.  
At the end the function $\Phi(\alpha)$ will be obtained as the saddle-point value of the following functional over the $n \times n$ diagonal block $Q$:
\beq
\Phi(\alpha,Q)=\alpha \beta F[Q]+S[Q,\alpha]
\label{functional}
\eeq
where $F[Q]$ is the standard SK free energy functional of a $n \times n$ matrix $Q_{ab}$ that is zero on the diagonal:
\beq
F[Q]=-{\beta\over 4}+{\beta \over 2 n}\sum_{a<b}^nQ_{ab}^2-{1 \over \beta n}\ln \sum_{\{s\}}\exp[ \beta^2 \sum_{a<b}Q_{ab}s_as_b ]\, ,
\eeq
and $S[Q,\alpha]$ is the contribution of the fluctuations. Now $ S [Q, \alpha]$ is small for small $\alpha$ and it should be treated as a perturbation: in order to obtain the first non trivial term we should 
compute $S [Q, \alpha]$ at the saddle point of $F[Q]$. This is not a easy task: indeed we find that  $S [Q, \alpha]$ can be written as:
\begin{widetext}
\beq
S[Q,\alpha]  =  -{\alpha^2 \over N^2}\ln \int \left(\prod_{i<j}^{-N/n}\prod_{ab}^{n}{dQ^{ij}_{ab} \over \sqrt{ 2 \pi}}\right) \exp\left[-{1\over 2}\sum_{i<j}^{ -N/n}\sum_{ab}^n(1-\beta^2 \lambda_a \lambda_b+p)(Q^{ij}_{ab})^2+\left(- \alpha \over N\right)^{{1 \over 2}} {\beta^3}\sum_{i<j<k}^{-N/n}\sum_{abc}^nQ^{ij}_{ab}Q^{jk}_{bc}Q^{ki}_{ca}\lambda_a\lambda_b\lambda_c \right]\, ,
\label{S}
\eeq
\end{widetext}
where $\lambda_a$ with $a=1,\dots,n$ are the eigenvalues of the matrix $P_{ab}\equiv \langle s_a s_b \rangle$ and the averages $\langle \cdot \rangle$ are computed with respect to a single diagonal block. In particular if $Q_{ab}$ extremizes $F[Q]$ we have $Q_{ab}=\langle s_a s_b \rangle$ and therefore $P_{ab}=\delta_{ab}+Q_{ab}$.
The above expression is valid at the third order in $Q^{ij}_{ab}$ which is enough to get the first non-linear term in $\Phi(\alpha)$.
The diagonal structure above has been obtained performing various manipulations on the original integral.
In particular the matrix $P$ enters only though its eigenvalues  because the relevant expressions  are rotationally invariant at the order considered and the integral in $S[Q,\alpha]$ is invariant under a simultaneous rotation of all the blocks $Q^{ij}$.  
The  parameter $p$ has been introduced for later convenience and has to be put to zero eventually (it has the physical meaning of adding  a small perturbation on the couplings of replicas in different blocks effectively removing the degeneracy of permutations among them). Note that in the previous expression the off-diagonal indices run from $1$ to $-N/n$ because we have exploited the fact that the function $S[Q,\alpha]$ in the thermodynamic limit is invariant under a rescaling $\{\alpha \rightarrow b \alpha,N\rightarrow b N\}$. 
 
The above expression is suitable for an expansion in powers of $\alpha$.  The first $O(\alpha^2)$ term is obtained performing the Gaussian integral:
\beq
S[Q,\alpha]={\alpha^2 \over 4 n^2}\sum_{ab}\ln (1-\beta^2 \lambda_a \lambda_b+p)+O(\alpha^3)
\label{gausscorr}
\eeq
In the limit $\alpha \rightarrow 0$ the extremum of $\Phi[Q,\alpha]$ is given by the extremum of $F[Q]$ with the first $O(\alpha^2)$ correction (\ref{gausscorr}) evaluated on the free solution where $P_{ab}=\delta_{ab}+Q_{ab}$.
In the $n \rightarrow 0$ limit the Gaussian correction turns out to be divergent if $p=0$, this is because the lowest eigenvalue $\lambda_0$ of $P_{ab}$ obeys the following relationship:  $\lambda_0 \equiv 1-\int_0^1q(x)dx=1/\beta$ that holds exactly at all temperatures \cite{S}.  
The divergence of the $O(\alpha^2)$ correction suggests that the first non-linear term in $\Phi(\alpha)$ has a power smaller than two. We also expect that the series in powers of $\alpha$ will be divergent at all orders and we will have to resum it in some way. 

Before going into the next step of the computation,   we consider a similar treatment of the spherical model. At zero temperature the energy is minus the largest eigenvalue of a random Gaussian $N \times N$ matrix. The logarithm of the probability that the largest eigenvalue is lower than its typical value is indeed $O(N^2)$ and has been recently computed \cite{DM}. These results tell us that in the zero-temperature limit we have: 
$\Phi(\alpha)=-\alpha\, \beta-\,{2 \over 3}|\,\alpha\, \beta|^{3 \over 2}+o(\alpha^{3/2})$.
Repeating the above procedure we obtain that for the spherical model $\Phi(\alpha)$ can be computed as the saddle-point value of a functional similar to (\ref{functional}). The first $O(\alpha)$ term depends on a symmetric $n \times n$ matrix $Q_{ab}$ (not vanishing on the diagonal) and on a parameter $z$ enforcing the spherical constraint:
\beq
F[Q,z]={\beta \over 4\,n}\sum_{ab}Q_{ab}^2-{z\over \beta}+{1 \over 2 \beta n}\Tr \ln[z I-{\beta^2 \over 2}Q]
\eeq  
where $I$ is the identity matrix.
The second term is equal (at the third order in $Q^{ij}_{ab}$) to expression (\ref{S}) provided we have (in matrix notation) $P=(2 z I-\beta^2 Q)^{-1}$.
The solution of the spherical model in the $\alpha \rightarrow 0$ limit is given by a replica-symmetric (RS) $n \times n$ matrix $Q_{ab} \equiv q=1-T$ \cite{KTJ} and again the lowest eigenvalue $\lambda_0$ of $P$ turns out to obey the relationship $\lambda_0=1/\beta$ leading to a divergent $O(\alpha^2)$ correction.

\begin{figure}[htb]
\begin{center}
\epsfig{file=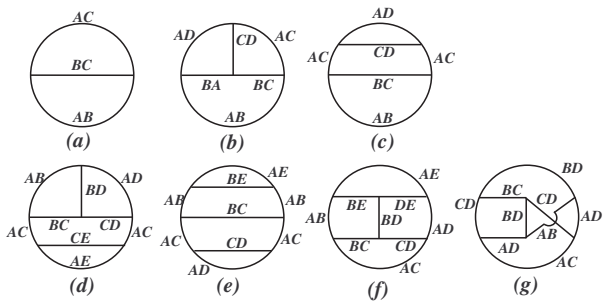,width=10cm}
\caption{Indexed Graphs  of the cubic theory up to four loops. Note that each index must belong to a closed path on the graph. The first six graphs are relevant because they satisfy the condition $K=L+1$ while graph $(g)$ is irrelevant because $K=4<L+1=5$ and therefore it yields an $O(N^{-1})$ contribution.}
\label{gcubic}
\end{center}\end{figure}

We now face the task of computing the leading corrections that diverges in the $p \to 0$ limit. This can be done by studying the diagrammatic loop expansion of $S[Q,\alpha]$.
We introduce capital indices that corresponds to a couple of upper and lower indices $A=(^i_a)$.
Given a  graph with vertices of degree three we have to associate to each line in it a couple of {\it different} capital indices in such a way that each vertex of the graph corresponds to a term in the action of eq. (\ref{S}).
This limits the number of $K$ free indices and for a graphs with $L$ loops one can prove that $K \leq L+1$.
Valid indexed graphs $G_I$ at four loops are represented in fig. \ref{gcubic}. 
For a graph $G_I$ with $V$ vertices we have a factor $-(-\alpha/N)^{V/2+2}/V!$ times a factor $\beta^2 \lambda_a \lambda_b /( p+1-\beta^2 \lambda_a \lambda_b)$ where $ab$ are the lower indices on each line.
We sum over all the free $K$ upper indices (they can take values $1,\dots,-N/n$) and vertex permutations of the graph. This yields a factor $V!(-N/n)^K$
that has to be divided by an appropriate symmetry factor $M(G_I)$ to avoid overcounting.
Relevant graphs are those that yield an $O(N^2)$ contribution, {\it i.e.} they satisfy the condition $K=V/2+2$ or equivalently $K=L+1$. Consistently it can be shown that no graph can yield a contribution greater than $O(N^2)$.
Summarizing the sum over the upper indices yields the factor:
\beq
{(-\alpha)^K(-1)^{K+1} \over M(G_I) n^K}\, ,
\label{sumN}
\eeq 
that multiplies the result coming from the sum over the lower indices that we analyze now.
In order to sum over the lower indices we have to take care of the non trivial structure of the propagator.
In particular given an indexed graph  each line with lower indices $ab$ in the graph  corresponds to a factor
$\beta^2 \lambda_a \lambda_b /( p+1-\beta^2 \lambda_a \lambda_b)$.
The resulting object has to be summed over the $K$ free indices (each running from $1$ to $n$) leading to:
\beq
{1 \over n^K}\sum_{a_1,\dots,a_K}^n\prod_{lines}{\beta^2 \lambda_{a_i} \lambda_{a_j} \over p+1-\beta^2 \lambda_{a_i} \lambda_{a_j}}
\label{linefac}
\eeq
where we have borrowed the factor $n^{-K}$ from (\ref{sumN}).
To perform this summation we must use some well know results on the eigenvalues of a replica symmetric matrix $Q$ and a hierarchical matrix $Q_{ab}$ characterized by a function $q(x)$ \cite{MP}.

We consider the first RS case: after some algebra we find that at the leading order in $p$ the final result at leading order in $p$ is:
\begin{equation}
{(\beta q)^K \over p^{K+I}}\left.\left({d^K \over dx_1 \dots dx_K}\prod_{lines}{1 \over 1-x_a-x_b}\right)\right|_{x=0}\, ,
\end{equation}
where $I$ is the number of lines of the graph.

In the full RSB case it turns out that the dominant divergent contribution comes from shape of the function $q$ around $x=0$ that we assume to be linear.
At the end of the day we obtain that at leading order in $p$ the contribution of the diagram is:
\beq
{(\beta \sqrt{T \dot{q}(0)})^K \over p^{{K \over 2}+I}}A_{x_1}\circ \dots \circ A_{x_K} \circ \prod_{lines}{1 \over 1+{1 \over 2}\,{x_a^2}+{1 \over 2}\,{x_b^2}}
\eeq
where we used the operator $A_{x}[g]\equiv-\int_0^\infty{1 \over x}{dg \over dx}dx$.

As we already noticed a relevant diagram has $K=L+1$ therefore in the cubic theory we have $I=3K-6$. 
Rescaling the variable $p$ respectively as $p=( -{ \alpha \beta \sqrt{T \dot{q}(0)} / z})^{2 \over 7}$, $p=( -{ \alpha \beta q /z})^{1 \over 4}$ 
and multiplying by the factors coming from (\ref{sumN}) we get an expression of $S[Q,\alpha]$ in terms of two functions $f_{RS}(z)$ and $f_{fRSB}(z)$. The loop expansion yields the series of the two functions in powers of $z$ .   
Taking the $z \rightarrow \infty$ limit we eventually obtain:
\beqa
S[Q,\alpha] & = & (-\alpha \beta \sqrt{T \dot{q}(0)})^{12/7}\, C_{fRSB}+\, o(-\alpha)^{12/7}  \nonumber
\label{SQA}
\\
S[Q,\alpha] & = & (-\alpha \beta q)^{3/2}\, C_{RS}+\, o(-\alpha)^{3/2} 
\label{SQARS}
\eeqa
where $C_{fRSB}  \equiv  \lim_{z \rightarrow \infty}z^{2 \over 7}f_{fRSB}(z)$ and $C_{RS}  \equiv \lim_{z \rightarrow \infty}z^{1 \over 2}f_{RS}(z)$. 
At fourth loop order the diagrams are shown in fig. (\ref{gcubic}) and lead to the following expressions:
\beqa
f_{RSB}(z) & = & -{\pi \over 8}+0.456\, z-3.278\, z^2+36.11\, z^3+O(x^4).
\nonumber
\\
f_{RS}(z)& = & -{1 \over 4}+{7 \over 6}\,z-19\, z^2+443\, z^3+O(z^4).
\nonumber
\eeqa
The zero-th order terms comes from a similar treatment of the Gaussian contribution (\ref{gausscorr}).
The problem of extracting the $z \rightarrow \infty$ behaviour of functions like $f_{RSB}(z)$ and $f_{RS}(z)$ from their expansions in powers of $z$ has already appeared in spin-glass literature \cite{PRS1} in a similar context and various resummation methods have been devised. Performing similar analysis on the above series we obtained the following estimates: $C_{RS}=-.68$, $C_{fRSB}=-.66$ with a $15\%$ error.
The estimate for $C_{RS}$ is in good agreement with the exact result $C_{RS}=-2/3$ quoted above.

\begin{figure}[t]
\begin{center}
\epsfig{file=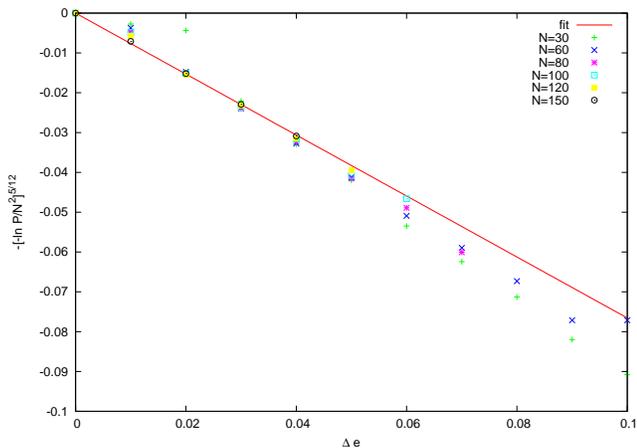,width=6cm,angle=-90}
\caption{Numerical Sample Complexity vs. Energy difference at zero temperature for the SK model (data of Ref. \cite{CMPP} courtesy of the Authors). The data are rescaled with the exponent $5/12$ and are well fitted by a linear function $-.765\,\Delta e$, which combined with the result $\lim_{T \rightarrow 0}T \dot{q}(0)=.743368$ leads to $C_{fRSB}=-.64(2)$}
\label{numdat}
\end{center}\end{figure}

In the $fRSB$ case we obtain from (\ref{SQA}):
\beq
{1 \over N^2}\ln P(\Delta f)=-a_{+}[T \, \dot{q}(0)]^{-6/5} \Delta f^{12/5}+o(\Delta f^{12/5})
\label{fCC}
\eeq
Where $a_{+} \equiv (-C_{fRSB})^{-7/5}\,35 (7/3)^{2/5}/ (2^{4/5} 144)$.
The numerical data of Ref. \cite{CMPP} at zero temperature are highly consistent with the $12/5$ scaling, see fig. (\ref{numdat}); a linear fit on the data corresponding to $N=150$ combined with $\lim_{T \rightarrow 0}T \dot{q}(0)=.743$ \cite{PR1} leads to an estimate $C_{fRSB}=-.64(2)$. 

Remarkably the previous results display a great deal of universality. Indeed the exponents $12/7$ and $3/2$ and the coefficients $C_{fRSB}$ and $C_{RS}$ depends only on the $fRSB$ or $RS$ structure of the eigenvalues of the matrix $P$.
The sole dependence on the actual model and on the temperature being respectively through the parameters $\sqrt{T \dot{q}(0)}$ and $q$. 
Interestingly enough the same universal behaviour is displayed by the function $L(\Delta f)$, again both for RS and fRSB \cite{PR1}.
For both  $L_2(\Delta f)$ and $L(\Delta f)$ we expect the universal behaviour to hold only at the leading order for small $\Delta f$, nevertheless this points towards universality of the corresponding small-deviations distribution. Indeed the reader may have already noticed that the $12/5$ exponent in (\ref{fCC}) is fully consistent with the assumptions that the large positive deviations match the far right tail of the distribution of the small-deviations of the free energy  density if they scales as $f-f_N=O(N^{1/6})$ \cite{CPSV}, where $f_N$ is the sample average at size $N$. At present the latter hypothesis is widely accepted in the literature after many recent numerical and theoretical investigations (see {\it e.g.} \cite{PR1,ABMM,CMPP}) and our findings add further support to it. 

We conjecture that the probability distribution $P(\delta)$ of the of the rescaled variable 
\beq
\delta =(f-f_N)N^{5/6} / \sqrt{T \dot{q}(0)}
\eeq
 is of the form 
$P(\delta)=\exp(-G(\delta))$
where the large $\delta $ behaviour of $G(\delta)$ matches with the behaviour near the origins of the functions $L_2(\Delta f)$ and $L(\Delta f)$, i.e.
\beqa
 G(\delta) & \simeq  & a_{-} | \delta |^{6/5}\ \ {\rm for }\ \delta \rightarrow -\infty
\\
 G(\delta) & \simeq  & a_{+} \delta ^{12/5}\ \ {\rm for }\ \delta \rightarrow +\infty
\eeqa
with $a_{-} =1.366$ \cite{PR1} and $a_+=.36$. 

It is natural to assume that for a large class fRSB spin-glass model not only the large $\delta$ behaviour of $G(\delta)$  is independent from of the model, but that the function $G(\delta)$ does not depend on the model for all $\delta$ and it is the same at all temperatures below the critical one.. A similar results should be valid for replica symmetric models where the equivalent function $H((f-f_N)N^{2/3}/q)$ is given by Tracy-Widom function \cite{TW}.

\end{document}